\documentclass[%
 aip,apl,showplace,
 amsmath,amssymb,
 reprint,%
]{revtex4-1}
\usepackage{graphicx}
\usepackage{dcolumn}
\usepackage{bm}
\begin{document}

\preprint{AIP/123-QED}

\title{Excitation of whispering gallery modes with a "point-and-play", fiber-based, optical nano-antenna}

\author{Jonathan M. Ward}
\affiliation{Light-Matter Interactions for Quantum Technologies Unit, Okinawa Institute of Science and Technology Graduate University, Onna, Okinawa 904-0495, Japan}
\author{Fuchuan Lei}%
\email{fuchuan.lei@oist.jp}
\affiliation{Light-Matter Interactions for Quantum Technologies Unit, Okinawa Institute of Science and Technology Graduate University, Onna, Okinawa 904-0495, Japan}%
\author{Stephy Vincent}%
\affiliation{Light-Matter Interactions for Quantum Technologies Unit, Okinawa Institute of Science and Technology Graduate University, Onna, Okinawa 904-0495, Japan}%
\author{Pooja Gupta}%
\affiliation{CSIR-Central Scientific Instruments Organisation, Chandigarh, Sec.-30C, India}%
\affiliation{Academy of Scientific and Innovative Research (AcSIR), Ghaziabad, India}
\author{Samir K Mondal}%
\affiliation{CSIR-Central Scientific Instruments Organisation, Chandigarh, Sec.-30C, India}%
\author{Jochen Fick}
\affiliation{Université Grenoble Alpes, CNRS, Grenoble INP, Institut Néel, 38000 Grenoble, France}
\author{S\'ile Nic Chormaic}%
 \affiliation{Light-Matter Interactions for Quantum Technologies Unit, Okinawa Institute of Science and Technology Graduate University, Onna, Okinawa 904-0495, Japan}%
 \affiliation{Université Grenoble Alpes, CNRS, Grenoble INP, Institut Néel, 38000 Grenoble, France}

\date{\today}

\begin{abstract}
We demonstrate the excitation and detection of whispering gallery modes in optical microresonators using a "point-and-play", fiber-based, optical nano-antenna. The coupling mechanism is based on cavity-enhanced Rayleigh scattering. Collected spectra exhibit  Lorentzian dips, Fano shapes, or Lorentzian peaks, with a coupling efficiency around 13\%. The spectra are characterized by the coupling gap,  polarization, and  fiber tip position. The coupling method is simple, low-cost and, most importantly, the \textit{Q}-factor can be maintained at $10^8$ over a wide coupling range, thereby making it  suitable for metrology, sensing, or cavity quantum electrodynamics (cQED) experiments. 
\end{abstract}



\maketitle
The past decades have seen intense research efforts for the development of whispering gallery resonators (WGRs) using different materials and geometries \cite{Ward2014,2018arXiv180500062, 2018arXiv180904878L}. The combination of the extremely high optical quality (\textit{Q}) factor and small mode volume makes WGRs especially suitable for studying lasing behavior \cite{wu2010ZBNA, he2013whispering, lei2017pump}, nonlinear optics  \cite{YangNLO, lin2017nonlinear}, quantum optics \cite{vernooy1998cavity}, optical information processing  \cite{lei2017bandpass}, optomechanics \cite{aspelmeyer2014cavity}, sensing \cite{2018arXiv180500062,ward2018nanoparticle}, fundamental   physics \cite{peng2014parity}, etc.
To take WGRs out of the research laboratory  and increase their functionality for real-world applications, it is critical that a simpler way to couple light into and out of the resonator be found. In general, free-space beams cannot be used to excite  high \textit{Q}-factor modes efficiently due to phase mismatching \cite{matsko2006optical} between the  beam and the cavity modes.  This limitation can be overcome by introducing cavity deformation,  allowing chaotic rays to assist in the coupling \cite{jiang2017chaos}. Alternatively, the phase-matching condition can be satisfied  using  evanescent-wave coupling, achieved via, for example, a prism \cite{braginsky1989quality}, a side-polished fiber \cite{ilchenko1999pigtailing}, a tapered fiber \cite{knight1997phase} or a grating-based fiber coupler \cite{Farnesi:14}.  High  coupling efficiency from the near-field interaction is achievable, but these each have their own advantages and disadvantages;  the prism method is limited by its bulky size, while  it is difficult to satisfy the phase-matching condition for side-polished fibers.  For this reason,  tapered optical fiber coupling  is widely used, as it provides a fiber-integrated solution to the coupling problem and near-ideal coupling is attainable \cite{knight1997phase}. There are also limitations with this technique, e.g. the length of the tapered regions cannot be arbitrarily short since the adiabatic condition for efficient mode propagation must be satisfied \cite{love1991tapered}; this leads to tapered fibers being fairly fragile and they easily deteriorate due to dust accumulation on their surface. Previous works have demonstrated packaging of the tapered fiber and microresonator into glue or low refractive index materials \cite{yan2011packaged,wang2016packaged,zhao2017raman} to overcome this problem;  however, the cavity \textit{Q}-factors  decrease by such modifications. Another possible drawback of such a packaging method is that the resulting device has two pigtails - one input and one output port - making it somewhat cumbersome. 

In recent years, there have been several attempts at injecting light into, and collecting light from, WGRs using single-ended waveguides, including on-chip \cite{shu2012perpendicular,endfire} and fiber-based configurations \cite{bai2018whispering,liu2018whispering}. The  waveguide considerably modifies the geometry of the system and high-\textit{Q} modes cannot be maintained (limited to about $10^5$). Very recently, a single-ended waveguide configuration was demonstrated by Shu \textit{et al. }\cite{shu2018scatterer}; they placed a small Rayleigh scatterer, acting  as a nano-antenna, on the surface of the microsphere.  Light was focused onto the antenna using a graded-index (GRIN) fiber-coupled lens and coupling between the free-space beam and the WGMs was shown.  The collected light was in the far-field and coupled back into the launching fiber.  Due to the Purcell effect, a coupling efficiency of 16.8\% was achieved. Even though feasible, this configuration relied on a commercial GRIN lens and tricky  alignment since the focused beam must overlap with the nanoparticle. Note that WGMs can be directly mapped using a fiber-based optical nano-antenna, such as a near-field probe \cite{knight1995mapping}. However, to date, there is no report on complete use of both illumination and collection modes using the fiber-based antenna for WGRs.

In this work, we propose and demonstrate an alternative WGR coupling scheme that is  easy to make, requires  little alignment, and is both convenient and stable. This is the first, single-ended, fiber-based optical nano-antenna that can be used to \textit{simultaneously} excite and collect light from the WGMs of a microresonator, making it very promising for optical sensing applications or  strong coupling with quantum emitters.

\begin{figure}[h]
\centering
\includegraphics[width=8cm, height=4.5cm]{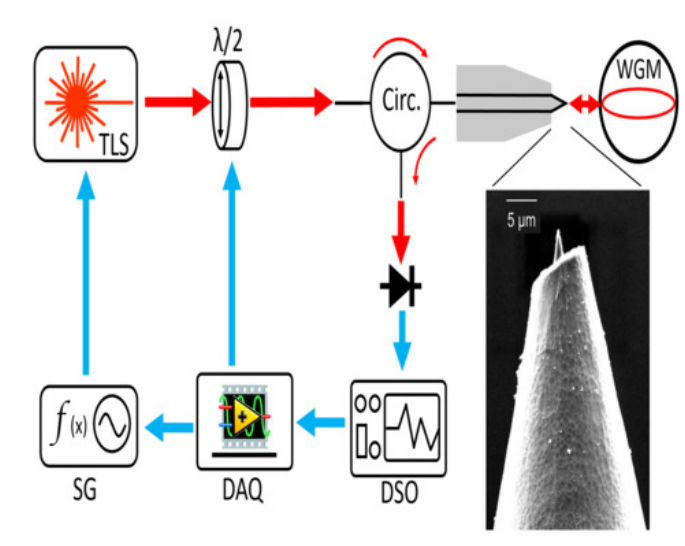}
\caption{Schematic of the coupling system. $\lambda$/2:Half wave plate; TLS: Tunable laser; Circ.: Optical circulator; DSO: Digital signal oscilloscope; SG: signal generator; DAQ: Digital Acquisition Card; WGM: whispering gallery mode. Inset: SEM image of the fiber-based nano-antenna.}
\label{fig1}
\end{figure}

Figure \ref{fig1} shows a schematic of the coupling system. A fiber-based optical antenna is brought into the evanescent field of a spherical WGR.  The inset of Fig. \ref{fig1} is a scanning electron microscope (SEM) image of the antenna, which was developed on the tip of a highly Germanium-doped single-mode optical fiber by etching in 48\%  hydrofluoric acid solution in a hydrophobic capillary tube. The etching process  exploits surface tension and capillary action, an interaction between contacting surfaces of a liquid and a solid, and different etching rates for the Germanium-doped optical fiber core and cladding materials. Full details of the fabrication   can be found in \cite{mondal2009optical,mondal2011ultrafine}. The end-face of the antenna has a curvature with a typical radius of 50 nm. The etching process produces two results that make this device unique. First, the etching does not alter the core diameter where protected by the cladding, only the cladding and exposed tip of the core are tapered. Second, the nano-antenna is formed directly above the core. These two features ensure good coupling between the antenna and the guided mode of the fiber, and eliminate the imprecise process of depositing a nanoparticle on the fiber tip. The microsphere resonator was fabricated from standard silica fiber using a $\rm {CO_2}$ laser \cite{lei2017bandpass}. The diameters of  microspheres  ranged from 50 $\mu$m to 200 $\mu$m.

\begin{figure}[t]
\centering
\includegraphics[width=0.8\linewidth]{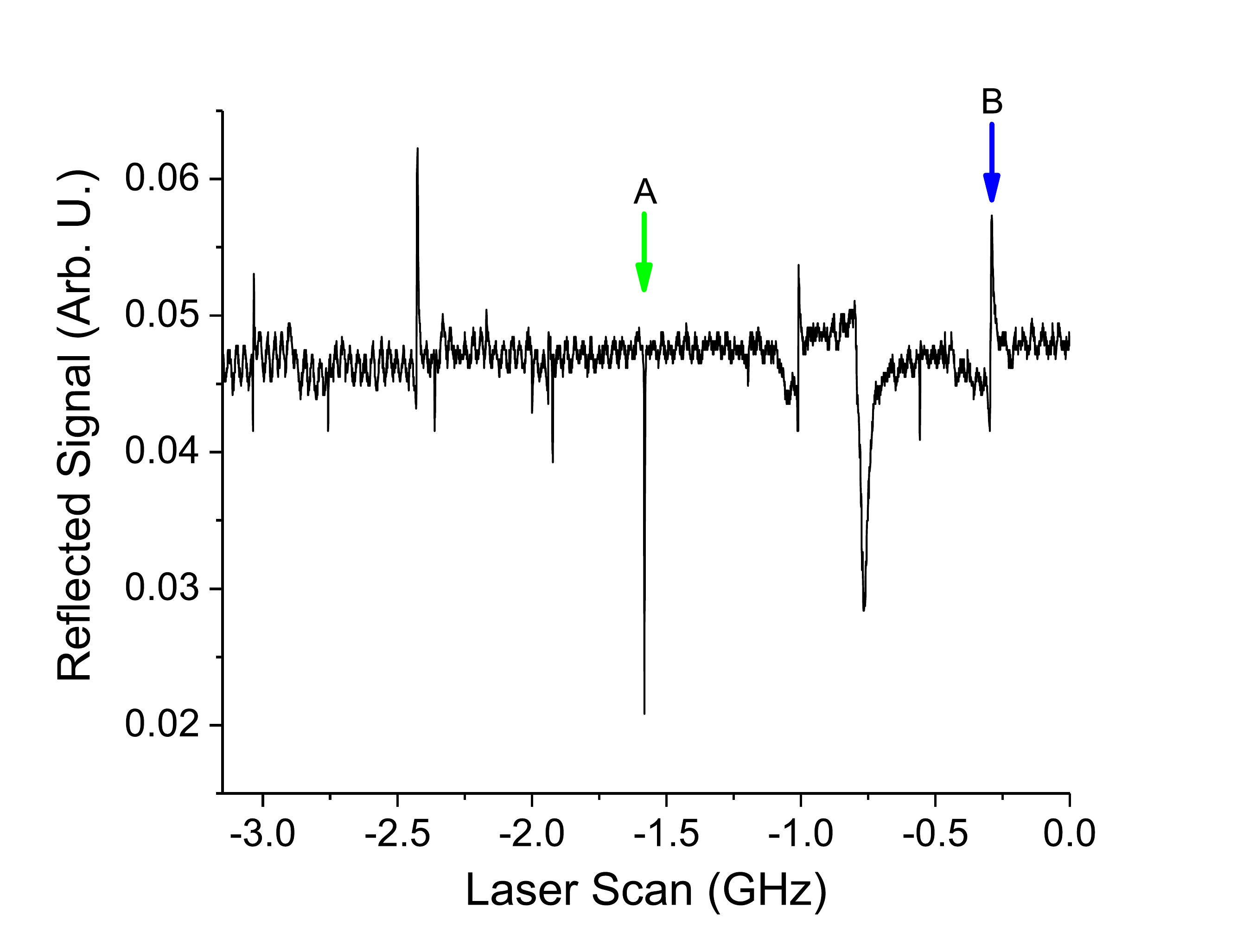}
\caption{A typical spectrum of the reflected signal when the fiber-based antenna is used for  excitation of resonances in a 186 $\mu$m diameter  microsphere. The two arrows indicate modes that were recorded as the gap was varied, see Fig. \ref{fig3}. }
\label{fig2}
\end{figure}

\begin{figure}[h]
\centering
\includegraphics[ width=1\linewidth,trim={0 0 0 0.75cm},clip]{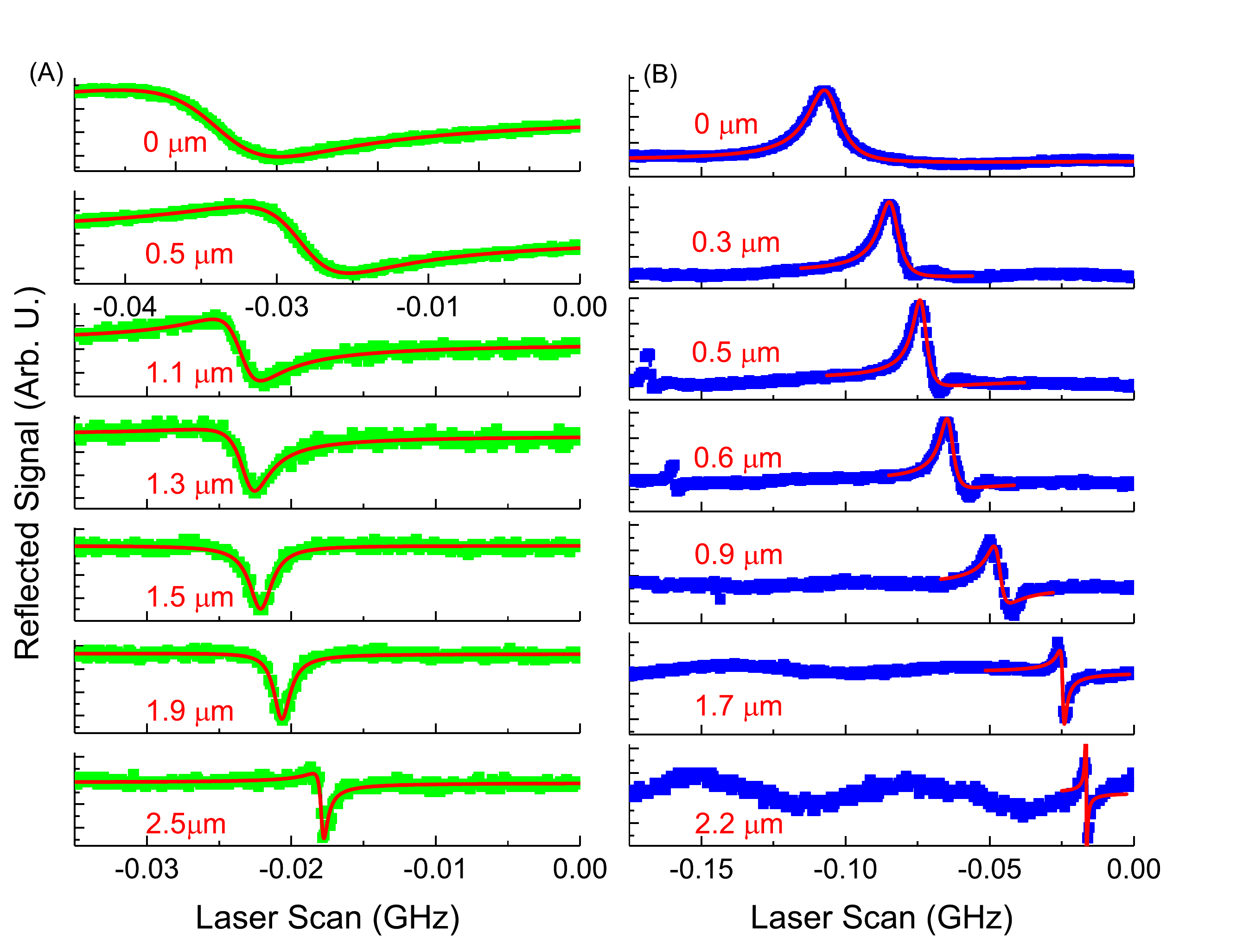}
\caption{Evolution of modes highlighted in Fig. \ref{fig2}. Top to bottom:  the gap between the antenna and microsphere increases from 0 to  2.2 $\mu$m. Solid red lines are fitted using Eqn. 1. The x-axis scale on the top two plots of (A) was adjusted to account for the broadened mode shape.}
\label{fig3}
\end{figure}

Light incident on the microsphere can excite WGMs that travel in both the clockwise (CW) and counterclockwise (CCW) directions via Rayleigh scattering from the antenna \cite{kippenberg2009purcell}. At the same time, some of the  light can be coupled back into the fiber pigtail through the antenna.
A tunable laser (Newport TLB  6700) in the C to L band wavelength range was used. A half-wave plate mounted on a motorized rotation stage controlled the polarization of the incident field prior to launching into the optical fiber. An in-line optical circulator (Circ.) facilitated both excitation of, and collection from, the WGMs. The reflected signal was monitored on a photodiode (PD). The antenna was kept fixed during experiments, whereas the movement of the microsphere was controlled using a three-axes piezo stage (Thorlabs NanoMax).  The microsphere was moved towards the antenna until a reflected signal was detected. A typical spectrum, see Fig. \ref{fig2}, contains not only Lorentzian dips, but also peaks and Fano resonances. An interesting feature is the change in the mode shape as a function of the gap between the antenna and the microsphere. The Lorentzian dip and Fano resonance highlighted in Fig. \ref{fig2} were monitored as the gap between the antenna and sphere was reduced to zero. The resulting spectra are plotted in Fig. \ref{fig3}. 

From top to bottom, the gap between the antenna and microsphere increases from 0 to approx. 2.2 $\mu$m. Note  there are three  effects occurring with a change in gap size: (i) the line shape of the cavity modes changes from a Fano shape to a Lorentzian peak and from a Lorentzian dip to a Fano shape, (ii) the base line changes, i.e, the amplitude of the reflected off-resonant signal varies with gap size, and (iii) a red-shift of  modes due to dispersion.  The first phenomenon can be  explained as an interference between two back-coupled fields through the fiber \cite{shu2012perpendicular}, one of which is a direct reflection from the surface of the microsphere, whereas the other is the field scattered from the cavity modes by the antenna. Hence, the total reflected field, $\rm E_{out}$, is composed of two parts such that 
\begin{eqnarray}
{\rm {E_{out}}}= re^{i\theta}{\rm E_{in}}+\frac{\kappa {\rm E_{in}}}{i(\omega-\omega_{0})-\gamma},
\end{eqnarray}
where $re^{i\theta}$ is the reflection coefficient, $\theta$ is the phase change between  input and output fields,  $\kappa$ is the coupling loss due to the antenna, $\omega_0$ is the resonant frequency, and $\gamma$ is the total decay ratio of the cavity mode, corresponding to a cavity  $Q=\omega/2\gamma$. Both the coupling loss and the reflection coefficient depend on the gap between the antenna and resonator. The collected field is negligible when the antenna is far  from the sphere since the evanescent wave diminishes exponentially with distance from the tip surface \cite{mondal2013evanescent}. Note that the field reflected from the surface of the sphere corresponds to a continuum spectrum, while the scattered cavity field corresponds to a discrete spectrum.  As a result,  Fano line shapes are formed \cite{limonov2017fano}.
\begin{figure}[h]
\centering
\includegraphics[width=0.8\linewidth]{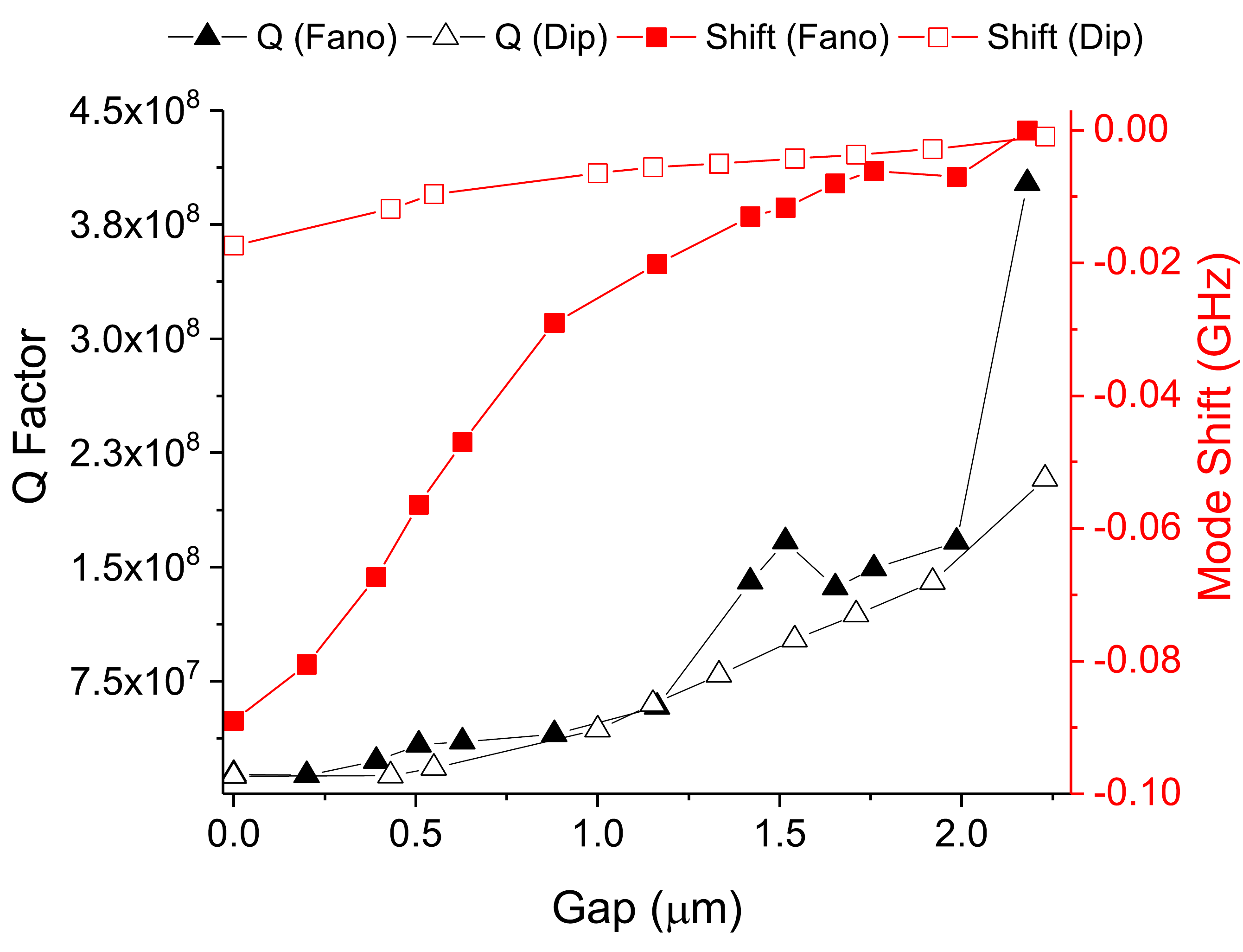}
\caption{The fitted \textit{Q}-factors and mode shifts for the modes in Fig. \ref{fig3}. The \textit{Q}-factors (mode shifts) of the Fano and  Lorentzian resonances are represented by  solid and empty triangles (squares), respectively.}
\label{fig4}
\end{figure}

 The antenna has both dispersive and dissipative effects on the modes, resulting in a mode shift and  broadening. By extracting $\gamma$ and the laser detuning from the fitted curves in Fig. \ref{fig3}, we can determine the corresponding  \textit{Q}-factor and mode shift as a function of  gap. These are plotted  in Fig. \ref{fig4}. For the modes, the shifts induced by the antenna are quite small, only 20 MHz for the resonant dip and 90 MHz for the Fano mode. As the Lorentzian dip changes to a Fano shape with decreasing gap (from 2.2 $\mu$m to zero), the linewidth increases from 0.8 MHz to 15 MHz. A similar linewidth change was observed as the Fano resonance changed to a Lorentzian peak. The effect of the antenna on the \textit{Q}-factor is comparable to that of a tapered optical fiber. Thus, it is possible to maintain a high-\textit{Q} with this "point-and-play" coupler,  unlike the previous end-fire coupler, which had a significant impact on the cavity \textit{Q}-factor \cite{endfire}.

\begin{figure}[h]
\centering
\includegraphics[width=0.8\linewidth]{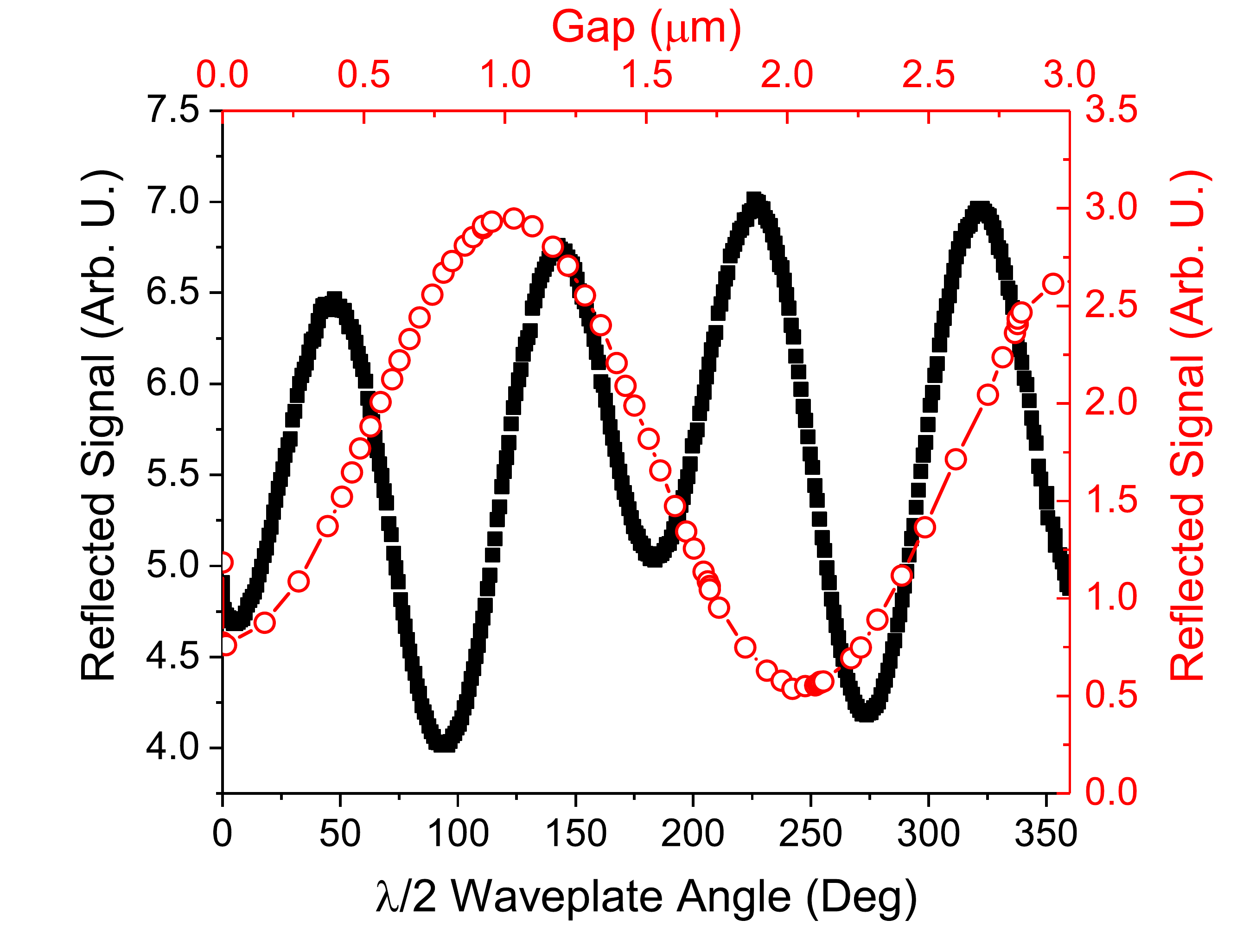}
\caption{The amplitude of the reflected off-resonant light as a function of gap (red circles and upper x-axis) and polarization (black squares and bottom x-axis).}
\label{fig5}
\end{figure}
\begin{figure}[h]
\centering
\includegraphics[width=1\linewidth,trim={0 0 0 1cm},clip]{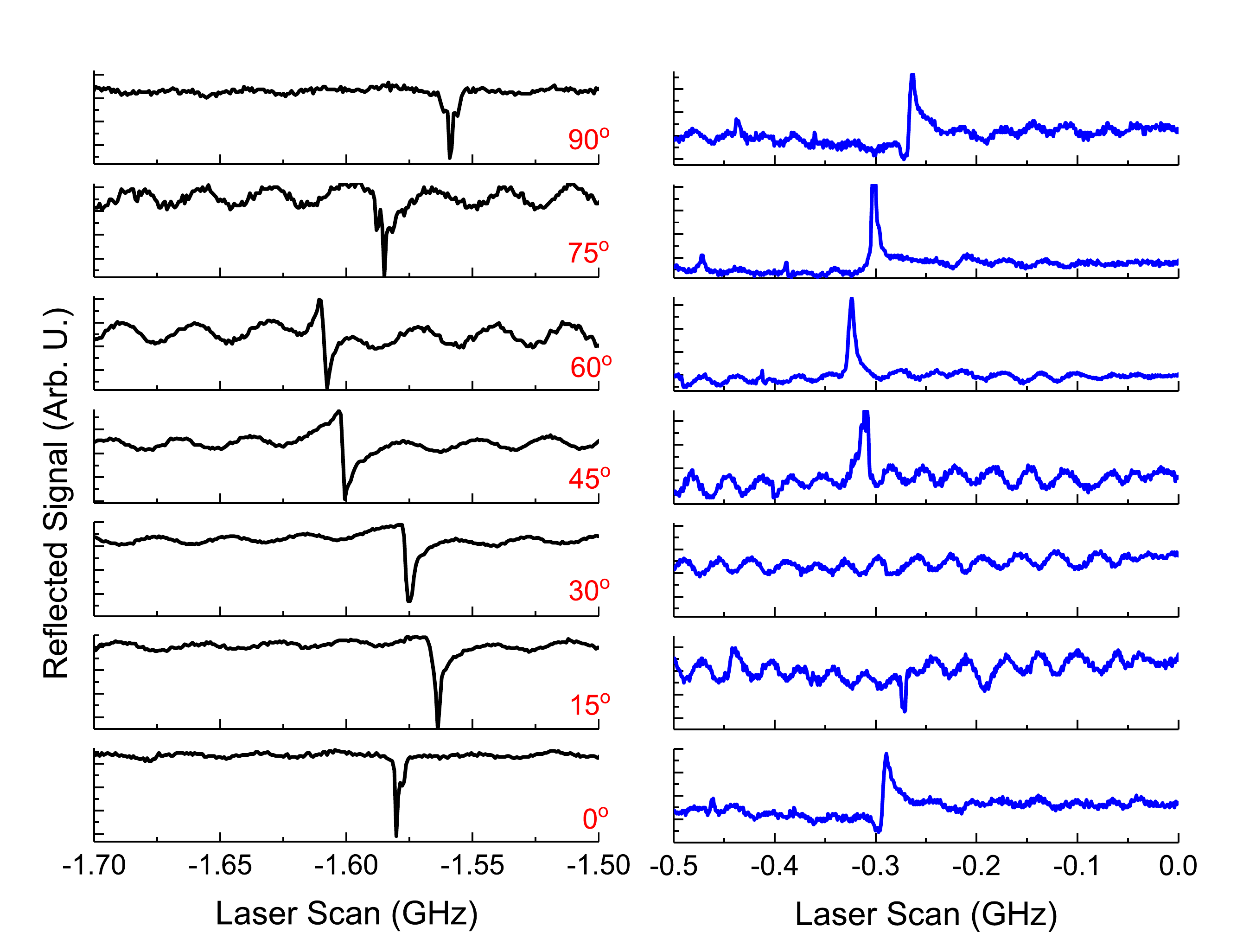}
\caption{Polarization dependent reflected spectra for a Lorentzian dip and Fano-shaped resonance. The polarization is changed by rotating the half-wave plate shown in Fig. \ref{fig1}.}
\label{fig6}
\end{figure}

The base line change with gap may also result from an interference  between the scattered and reflected light fields. From the samples under test, we found that the reflected off-resonant signal had a sinusoidal dependence with gap, see Fig. \ref{fig5}.  Besides the spectral dependence on the gap,  the amplitude of the reflected off-resonant light and the mode spectrum also depend on the input polarization. The polarization angle of the light launched into the fiber was used to strongly modify the mode shape and baseline transmission. A half-wave plate was used to rotate the polarization input to the optical fiber. The off-resonant amplitude and polarization dependent spectra are shown in Fig. \ref{fig5} and Fig. \ref{fig6}, respectively. These phenomena can be understood if we consider the following two points: (i) the fabricated antenna is not rotationally  symmetric and the transmittance through the antenna depends on the polarization,  and (ii) the excitation of WGMs depends on the polarization of the excitation light. In general, we can assume that the angular momentum of light from the nano-antenna includes both longitudinal  and transverse spin components, and may even contain some orbital angular momentum \cite{sarkar2014optical}. However, it is not easy to measure these components of light, particularly in the near-field as would be required. Nonetheless, WGM spectra were recorded for different polarization angles for two different modes, see Fig. \ref{fig6}. A Lorentzian dip transforms to a Fano every 45 degrees and then back to a dip every 90 degrees, whereas modes that were initially Fano convert to a Lorentzian peak every 45 degrees and then back to Fano every 90 degrees. Changing the polarization changes the coupling rate, the reflection coefficient, and excites modes with slightly different \textit{Q}-factors and wavelengths, thus giving rise to the variety of modes which can be recreated by Eq. 1. By optimizing the polarization and the coupling gap, the power of the highest measured peak, i.e., the maximum peak in Fig. \ref{fig6}, was 1.8\% of the input power. Assuming that the input coupling equals the output coupling, the maximum coupling efficiency to the WGMs should, in principle, exceed 13\%. 

\begin{figure}[h]
\centering
\includegraphics[width=9cm,height=7cm]{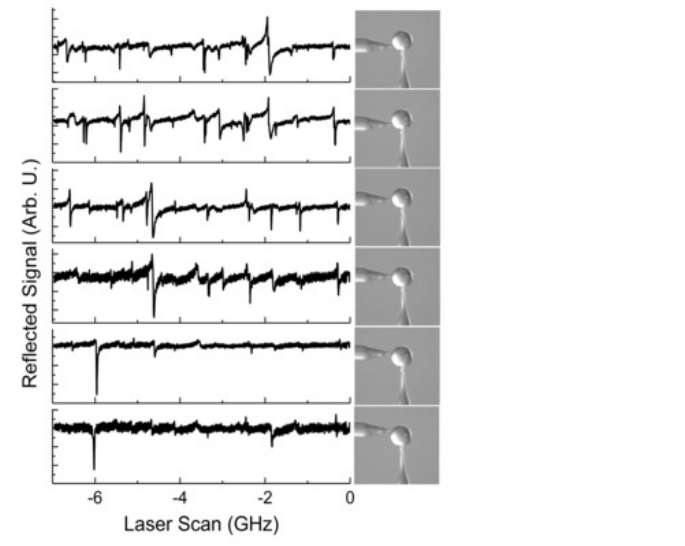}
\caption{Spectra for different polar positions of the fiber tip.}
\label{fig7}
\end{figure}

Finally, if one considers that the microsphere resonator can support different azimuthal modes and that Rayleigh scattering need not satisfy the phase-matching condition, it should be possible to achieve coupling over almost the entire surface of the resonator except for a small region around the support stem. To confirm this experimentally, the reflected spectrum was recorded for different polar positions on the sphere, see Fig. \ref{fig7}. The fiber tip was placed into contact with the sphere at each point, whereas neither the polarization nor angle of the fiber were controlled. At each position it was possible to excite WGMs and there was even some consistency between successive points, implying the coupler should be relatively robust against mechanical vibrations.  This could prove to be particularly convenient when the system is in a noisy environment, where a standard tapered fiber coupler may not be feasible. 
As a final point, it was also noted that the fiber tip did not degrade over time (at least one week) unlike tapered optical fibers which are exceptionally prone to dust accumulation.

Based on our results, we propose that the fiber-based optical antenna coupling method is suitable for building an ultracompact WGR-based sensor. There are a number of improvements that could be explored, e.g., a more symmetric tip, a specific tip design via FIB milling, or metallic coatings to provide plasmonic enhancement. Geometric considerations such as the tip height, tip base diameter, fiber core size, and core/tip overlap could also be studied to improve coupling efficiency. Ultimately, the microresonator and the antenna could be packaged together as a minuscule probe, with a lateral size of 125 $\mu$m, depending on the fiber. Such probes hold promise for producing low-cost, portable, and highly sensitive point-of-care (POC) diagnostic devices.

\section*{Funding }
This work was partly funded by the Okinawa Institute of Science and Technology (OIST) Graduate University.
\bibliography{reff}

\end{document}